\documentclass[
    ,final            % use final for the camera ready runs
  ]
  {aipproc}

\usepackage{amsmath}
\usepackage{txfonts}

\layoutstyle{6x9}

\newcommand{\rmi}{\mathrm{i}}
\newcommand{\rmd}{\mathrm{d}}
\newcommand{\GF}{G_\mathrm{F}}
\newcommand{\vs}{\vec{s}}
\newcommand{\vez}{\vec{e}_z}
\newcommand{\vH}{\vec{H}}
\newcommand{\vtH}{\vec{\tilde{H}}}
\newcommand{\wpr}{\omega_\mathrm{pr}}

\begin{document}

\title{A simple model for spectral swapping of supernova neutrinos}
%\preprint{INT-PUB-09-033}
\classification{14.60.Pq, 97.60Bw}
\keywords      {supernovae, neutrino oscillations}

\author{Huaiyu Duan}{
  address={Institute for Nuclear Theory,
University of Washington, Seattle, WA 98195, USA}
}

\begin{abstract}
Neutrinos emitted from a core-collapse supernova can experience
collective flavor transformation because of high neutrino fluxes. 
As a result, neutrinos of different flavors can have their
energy spectra (partially) swapped, a phenomenon known as the (stepwise)
spectral swapping or spectral split. We give a brief review of a
simple model that explains this phenomenon. 
\end{abstract}

\maketitle

%%%%%%%%%%%%%%%%%%%%%%%%%%%%%%%%%%%%%%%%%%%%
%% MAINMATTER
%%%%%%%%%%%%%%%%%%%%%%%%%%%%%%%%%%%%%%%%%%%%

\section{Introduction}

\begin{figure}
\includegraphics[width=.5\textwidth]{spec-inu}
\includegraphics[width=.5\textwidth]{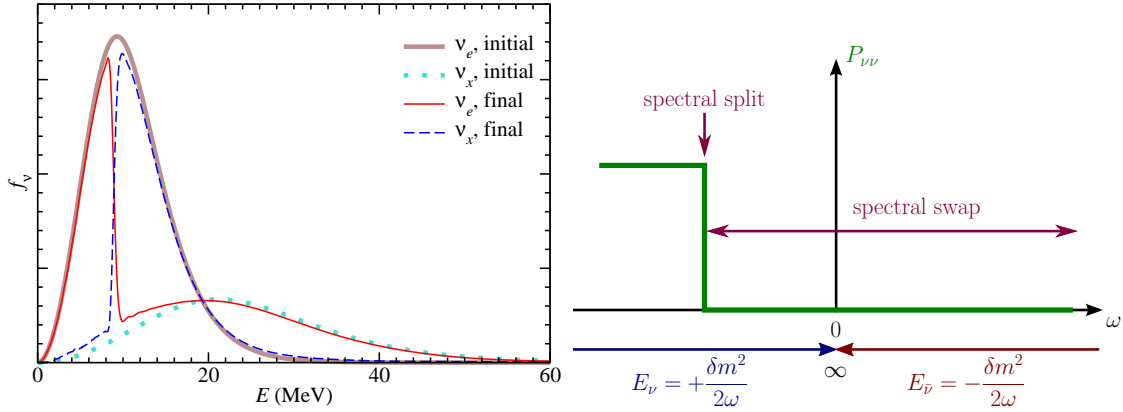}
  \caption{\label{fig:swap}The initial and final energy spectra of
  $\nu_e$ and $\nu_x$ (left panel) and the corresponding 
  (schematic) neutrino
  survival probability $P_{\nu\nu}$ (right panel)
in one of the numerical simulations
  (single-angle approximation, two-flavor mixing and inverted neutrino
  mass hierarchy) in \cite{Duan:2006an}.}
\end{figure}

During a core-collapse supernova event, the neutrino fluxes near the
nascent neutron star can be so large that the change of neutrino
refractive indices due to neutrino-neutrino forward scattering or
neutrino self-interaction becomes important
\cite{Fuller:1987aa,Notzold:1988kx}.
It was demonstrated in several numerical simulations 
(e.g., \cite{Duan:2006an,Duan:2006jv,Fogli:2007bk,Dasgupta:2007ws})
that neutrinos of different flavors can swap (parts of) their energy
spectra as they traverse the supernova envelope (see
Figure \ref{fig:swap}). Here we briefly reviewed a simple model 
that explains this spectral swap/split phenomenon.
(See \cite{Duan:2009cd} for a more detailed review.)

With neutrino self-interaction neutrinos with various energies and
propagating in different directions can be coupled together, which
makes it difficult to analyze neutrino flavor transformation in supernovae.
For simplicity we adopt a spherically symmetric supernova model under
the ``single-angle approximation''. The idea described in the rest of
the talk can be generalized to anisotropic environments
\cite{Duan:2008fd}. In this simple model we assume that the flavor
evolution of neutrinos is independent of neutrino propagation
directions. The flavor state $|\psi_{\nu,E}(r)\rangle$ of a
neutrino $\nu$ with energy $E$ at radius $r$ can be solved from the
Schr\"odinger-like equation
\begin{equation}
\rmi\frac{\rmd}{\rmd r}|\psi_{\nu,E}(r)\rangle=
\hat{H} |\psi_{\nu,E}(r)\rangle.
\label{eq:eom0}
\end{equation}
In the flavor basis
the Hamilton in equation \eqref{eq:eom0} can be written
as \cite{Qian:1994wh,Pantaleone:1992xh}
\begin{equation}
\mathsf{H}=\frac{\mathsf{M}^2}{2E}
+\sqrt{2}\GF\,\mathrm{diag}[n_e(r),0,0]
+\sqrt{2}\GF\int_0^\infty\rmd E^\prime\, 
[\rhoup_{E^\prime}(r)-\bar\rhoup_{E^\prime}(r)],
\label{eq:H}
\end{equation}
where $\mathsf{M}$ is the neutrino mass matrix, $\GF$ is the Fermi's
constant, $n_e(r)$ is the electron number density, and $\rhoup_{E^\prime}(r)$
and $\bar\rhoup_{E^\prime}(r)$ are the flavor density matrices for
the neutrino and the antineutrino with energy $E^\prime$, respectively
\cite{Sigl:1992fn}.

\section{Neutrino flavor isospin}

We shall use the two-flavor mixing scheme $(\nu_e,\nu_x)$ where
$\nu_x$ is an appropriate mixture of $\nu_\mu$ and $\nu_\tau$. We take
the effective vacuum mixing angle and the mass-squared difference to be
$\theta_\mathrm{v}\approx\theta_{13}\ll1$ and $\delta m^2\approx\pm\delta
m^2_\mathrm{atm}$, respectively,
where the plus (minus) sign is for the normal
(inverted) neutrino mass hierarchy. The two-flavor neutrino
oscillation can be visualized with the help of, e.g., the neutrino
flavor isospin (NFIS) \cite{Duan:2005cp}
\begin{equation}
\vs_\omega=\left\{\begin{array}{lll}
\psi_{\nu,E}^\dagger(\vec{\sigma}/2)\psi_{\nu,E},
&\text{ for neutrinos }
(\omega=\delta m^2/2E),\\
(\sigma_y\psi_{\bar\nu,E})^\dagger(\vec{\sigma}/2)(\sigma_y\psi_{\bar\nu,E}),
&\text{ for antineutrinos }
(\omega=-\delta m^2/2E).
\end{array}\right.
\label{eq:s}
\end{equation}
Equations \eqref{eq:eom0} and \eqref{eq:H} now become
\begin{equation}
\frac{\rmd}{\rmd r}\vs_\omega
=\vs_\omega\times\vH_\omega
=\vs_\omega\times(\omega\vez^\mathrm{v}
-\sqrt{2}\GF n_e\vez^\mathrm{f}
-\mu\langle\vs\rangle).
\label{eq:eom}
\end{equation}
In equation \eqref{eq:eom} $\vez^\mathrm{v}$ and $\vez^\mathrm{f}$ are the unit
vectors in flavor space along which NFIS's representing
$|\nu_1\rangle$ and $|\nu_e\rangle$ must be oriented, respectively. The
misalignment of $\vez^\mathrm{v}$ and $\vez^\mathrm{f}$ (with
$2\theta_\mathrm{v}$ between them) signifies the non-coincidence of the
mass basis and the flavor basis of neutrinos. 
In vacuum an NFIS $\vs_\omega$ initially aligned with
$\vez^\mathrm{f}$ (i.e., a pure $\nu_e$)
precesses about $\vez^\mathrm{v}$ with (angular) frequency
$\omega$. When projected onto the $\vez^\mathrm{f}$-axis, this
precession yields vacuum oscillations with 
$|\langle\nu_e|\psi_{\nu,E}\rangle|^2=\frac{1}{2}+\vs_\omega\cdot\vez^\mathrm{f}$.
We emphasize that NFIS $\vs_\omega$ precesses in \emph{flavor space}
not coordinate space and has nothing to do with the internal spin of
the neutrino.
Also in equation \eqref{eq:eom} $\mu=2\sqrt{2}\GF n_\nu^\mathrm{tot}$ is the
coupling strength between NFIS's, and 
$\langle\vs\rangle=\int_{-\infty}^\infty\rmd\omega^\prime
f_{\omega^\prime}\vs_{\omega^\prime}$ is the average NFIS, where
$n_\nu^\mathrm{tot}$ is the total neutrino number density and
$f_{\omega^\prime}$ is the (constant) normalized distribution function
determined by neutrino energy spectra at the neutrino sphere.

The effects of neutrino masses, ordinary matter and background neutrinos
on collective neutrino oscillations can be easily gleaned
from equation \eqref{eq:eom}: (a) Because the mass term [the first term of
$\vH_\omega$ in equation \eqref{eq:eom}] is different for
neutrinos/antineutrinos with different energies, it induces different
precession of the corresponding NFIS's. In other words, the mass term
tends to disrupt the collective motion of NFIS's or the 
collective neutrino oscillation. (b) Because the matter term
(the second term of $\vH_\omega$) is independent of $\omega$, it is
neutral to collective neutrino oscillations and can be ``ignored''.%
\footnote{The effect of ordinary matter is subtle. When $n_e$ is
large, the mass basis is essentially replaced by the flavor basis with
a small effective mixing angle no matter what the original value of
$\theta_\mathrm{v}$
is \cite{Duan:2005cp,Hannestad:2006nj,Duan:2008za}. A large matter
density can also suppress collective neutrino oscillations in
anisotropic environments
\cite{EstebanPretel:2008ni,Duan:2008fd}.}
(c) The last term of $\vH_\omega$, representing neutrino
self-interaction, couples NFIS's with different values of $\omega$
and, therefore,  
drives collective neutrino oscillations.

\section{collective precession mode}

For simplicity we ignore the matter effect for the reason mentioned above.
The equations of motion \eqref{eq:eom} for NFIS's possess a
symmetry about $\vez$. (From now on we do not distinguish between
$\vez^\mathrm{f}$ and $\vez^\mathrm{v}$ because $\theta_\mathrm{v}\ll1$.)
If a group of NFIS's $\{\vs_\omega|\omega\in(-\infty,+\infty)\}$
satisfies equation \eqref{eq:eom}, then
$\{\vec{\tilde{s}}_\omega|\omega\in(-\infty,+\infty)\}$ also obey
equation \eqref{eq:eom}, where $\vec{\tilde{s}}_\omega$ are obtained by
rotating $\vs_\omega$ about $\vez$ by an $\omega$-independent angle
$\phi$. This symmetry
suggests the existence of a collective neutrino oscillation mode
in which all the NFIS's precess about $\vez$ with the same angular
frequency $\wpr$ \cite{Duan:2007mv,Duan:2007fw,Duan:2008fd}. In
this collective precession mode and with constant $n_\nu^\mathrm{tot}$
all NFIS's must be static in the reference frame which rotates about
$\vez$ with $\wpr$. It follows that  $\vs_\omega$ must be either aligned
or antialigned with $\vtH_\omega=\vH_\omega-\wpr\vez$, the effective
field experience by $\vs_\omega$ in this corotating
frame \cite{Duan:2006an}, and 
\begin{equation}
\vs_\omega=\frac{\epsilon_\omega}{2}\frac{\vtH_\omega}{|\vtH_\omega|},
\quad\epsilon_\omega=\pm1.
\end{equation}
One can
then show that all NFIS's in a collective precession mode is fully
determined by $\langle s_z\rangle$ 
and $\langle s_\perp\rangle$, the 
components of $\langle\vs\rangle$ that are parallel and orthogonal to
$\vez$, respectively, and the common precession frequency $\wpr$. The value of
$\langle s_z\rangle$ is fixed by the conservation law that arises from the
symmetry described above \cite{Hannestad:2006nj}, and $\wpr$ and
$\langle s_\perp\rangle$ can be solved from \cite{Raffelt:2007cb}
\begin{align}
1&=-\frac{1}{2}\int_{-\infty}^\infty\rmd\omega
\frac{\epsilon_\omega f_\omega}%
{\sqrt{[(\omega-\wpr)/\mu-\langle s_z\rangle]^2+\langle s_\perp\rangle^2}},\\
\wpr&=-\frac{1}{2}\int_{-\infty}^\infty\rmd\omega
\frac{\omega\epsilon_\omega f_\omega}%
{\sqrt{[(\omega-\wpr)/\mu-\langle s_z\rangle]^2+\langle s_\perp\rangle^2}}.
\end{align} 

If $n_\nu^\mathrm{tot}$ decreases slowly with $r$, neutrinos will stay
in the collective precession mode, and $\vs_\omega$ remains either
aligned or antialigned with $\vtH_\omega$. Note that this is similar
to the adiabatic Mikheyev-Smirnov-Wolfenstein (MSW) flavor
transformation \cite{Wolfenstein:1977ue,Mikheyev:1985aa}. 
In the adiabatic MSW transformation a neutrino stays
in either of the two eigenstates of $\mathsf{H}_\omega$ (with
non-vanishing $n_e$ but zero neutrino flux) which are the two NFIS
states aligned and antialigned with $\vH_\omega$, respectively. If
the collective precession mode of neutrino oscillations does not break
down until $n_\nu^\mathrm{tot}$ becomes negligible, all the NFIS's
will become either aligned or antialigned with $\vez$, and
\begin{equation}
\vs_\omega\cdot\vez=\frac{\epsilon_\omega}{2}\,\mathrm{sgn}(\omega-\wpr^0)=
\left\{\begin{array}{ll}
|\langle\nu_e|\psi_{\nu,E}\rangle|^2-\frac{1}{2}
&\text{for neutrino,}\\
\frac{1}{2}-|\langle\bar\nu_e|\psi_{\bar\nu,E}\rangle|^2
&\text{for antineutrino,}
\end{array}\right.
\end{equation} 
where $\wpr^0=\wpr|_{n_\nu^\mathrm{tot}\rightarrow0}$. Correspondingly,
$P_{\nu\nu}$, the probability for the neutrino/antineutrino to be in
the same flavor as it is at the neutrino sphere, is a step function
of $\omega$ (the right panel of Figure \ref{fig:swap}). 
This is exactly what is observed in numerical simulations (left panel
of Figure \ref{fig:swap}). This phenomenon is called ``stepwise
spectral swapping'' because $\nu_e$ and $\nu_x$ swap energy spectra at
energies above (below) the critical energy $E_\mathrm{C}=|\delta
m^2/2\wpr^0|$ in an inverted (normal) mass hierarchy
case \cite{Duan:2006an}. This is also known as the ``spectral split''
because $E_\mathrm{C}$ ``splits the transformed spectrum sharply into parts
of almost pure but different flavors'' \cite{Raffelt:2007cb}.

%%%%%%%%%%%%%%%%%%%%%%%%%%%%%%%%%%%%%%%%%%%%%%%%
%% BACKMATTER
%%%%%%%%%%%%%%%%%%%%%%%%%%%%%%%%%%%%%%%%%%%%%%%%

\begin{theacknowledgments}
The author would like to thank the conference organizers for the
opportunity to give this talk. The author also thank J.~Carlson,
G.~M.~Fuller and Y.-Z.~Qian for useful discussions. This work is
supported in part by US DOE Grant No. DE-FG02-00ER41132 at INT.
\end{theacknowledgments}



\bibliographystyle{aipproc}   % if natbib is available

\bibliography{refs}

\end{document}